\documentclass[twocolumn,aps,prc,showpacs,10pt,nofootinbib]{revtex4}

\renewcommand{\vec}[1]{\mbox{\boldmath $#1$}}

\usepackage{amsmath}
\usepackage{amssymb}
\usepackage{graphicx}
\usepackage{mathrsfs}

\begin{document}
\title{Reaction cross sections of hypernuclei and the shrinkage effect}

\author{T. Akaishi}
\affiliation{Department of Physics, Tohoku University, Sendai 980-8578, Japan}
\author{K. Hagino}
\affiliation{Department of Physics, Tohoku University, Sendai 980-8578, Japan}

\begin{abstract}
We calculate the reaction cross sections for $^6{\rm Li}$ and 
$^7_{\Lambda}{\rm Li}$
on a $^{12}{\rm C}$ target at $100\,{\rm MeV/nucleon}$ using the Glauber theory.
To this end, we assume a two-body cluster structure for $^6$Li 
and 
$^7_{\Lambda}{\rm Li}$, and employ the few-body treatment of the 
Glauber theory, that is beyond the well known optical limit approximation. 
We show that the reaction 
cross section for $^7_{\Lambda}{\rm Li}$ is 
smaller than that for $^6$Li by about 4\%, reflecting the shrinkage 
effect of the $\Lambda$ particle. 
\end{abstract}


\maketitle

One of the main interests in hypernuclear physics is the change of 
the nuclear structure due to a few $\Lambda$ particles, which is 
referred to as the impurity effect
\cite{motoba,motoba2,hiyama,HT06,tanida,HagiKoi,myaing,myaing2, 
Yao11,Hagino13,Minato13,Minato12,Minato09,Isaka,drip,drip2}. 
A $\Lambda$ particle can enter the interior of a nucleus since 
it does not receive the Pauli exclusion principle from nucleons, 
and attracts the surrounding nucleons. 
Various theoretical analyses and experimental measurements have suggested that this effect appears as a shrinkage of a nucleus, that is, 
the change of the nuclear size 
\cite{motoba,motoba2,hiyama,HT06,tanida,HagiKoi}. 
Particularly, a large shrinkage effect is expected 
for light nuclei which have cluster structures
\cite{motoba,motoba2,hiyama}, since clusters are in general 
weakly 
bound in these nuclei. 
The shrinkage of hypernuclei has been investigated 
experimentally 
with the $\gamma$-ray spectroscopy\cite{HT06,tanida}.
In the experiment of Ref.\cite{tanida}, 
the electric quadrupole transition probability $B(E2)$
from the excited $5/2^+$ state 
to the ground state in $^7_{\Lambda}{\rm Li}$  was measured.
The observed $B(E2)$ value of $^7_{\Lambda}{\rm Li}$ was smaller than 
that of $^6{\rm Li}$ by about $33\%$.
This corresponds to 
about $19\%$ shrinkage of the intercluster distance 
if one assumes the two-body cluster structure with 
core+deuteron (that is, $\alpha+d$ and $^5_\Lambda$He$+d$ for 
$^6$Li and $^7_\Lambda$Li, respectively)\cite{tanida}. 

In this paper, we investigate the shrinkage effect 
of $\Lambda$ particle using reaction cross sections
with the Glauber theory.
The reaction cross section is defined as 
a sum of all cross sections except for the elastic scattering, 
and it has played an important role in 
the discussion of a density distribution for 
neutron-rich nuclei, such as a ``halo" structure of exotic nuclei 
\cite{tanihata,Ozawa,tanihata13,hagino13,takechi}.
Classically, 
if a projectile and a target nuclei are assumed to be spheres with 
a radius of $R_P$ and $R_T$, respectively, 
the reaction cross section $\sigma_{\rm R}$ is given as 
$\sigma_{\mathrm R}=\pi(R_P+R_T)^2$. 
The experimental cross sections have indeed shown that 
the reaction cross section increases for nuclei which have a halo 
structure 
\footnote{Experimentally, the interaction cross sections, defined as 
a sum of cross sections in which the nucleon number changes, are 
actually measured instead of 
the reaction cross sections 
at intermediate and high energies. 
At these energies the interaction and the reaction 
cross sections do not differ much, 
especially for weakly bound systems,  
because inelastic scattering cross sections are small.}, 
and the reaction cross sections have thus been used as 
a standard tool to 
extract the nuclear size of neutron-rich nuclei\cite{Ozawa}.
It is thus of intriguing to study the reaction cross sections 
of hypernuclei and discuss their size. 

In order to extract the nuclear size and matter distributions,
the Glauber theory\cite{glauber} has often been 
used\cite{bertsch,AN03,Ogawa92,Ogawa01}.
Notice that the Glauber-type analyses need only the ground state information. 
This makes our study 
complementary to the method with electromagnetic transitions, which 
involves both the ground and excited states. 
 
We choose $^6{\rm Li}$ and $^7_{\Lambda}{\rm Li}$ nuclei as projectiles 
and $^{12}{\rm C}$ as a target.
Since these projectile nuclei are known to have a two-body
cluster structure 
\cite{HagiKoi,merchant}, we adopt the semi-microscopic cluster model 
\cite{merchant,buck} 
in order to obtain the ground state wave functions. 
In this model, 
the intercluster potentials between the core and deuteron are 
constructed based on the core+p+n structure. 
For the $s$-wave state (that is, the ground state), 
it is given as, 
\begin{eqnarray}
\begin{aligned}
V(R)=\int d\vec{r} 
\left[V_{\rm cN}(\vec{R}+\vec{r}/2)+V_{\rm cN}(\vec{R}-\vec{r}/2)\right]
\cdot|\psi_d(\vec{r})|^2, 
\end{aligned}
\label{pot}
\end{eqnarray}
where $\psi_d(\vec{r})$ is an $s$-state wave function for the 
relative motion between the proton and the neutron 
in the deuteron cluster. 
$V_{\rm cN}$ is 
the potential between the nucleons and the core nucleus. 
In our calculations, we employ an exponential form for 
the deuteron wave function\cite{sack}:
$\psi_d(r)=\sqrt{2\alpha}e^{-\alpha r}/\sqrt{4\pi}r$, 
$\alpha=0.2316\, {\rm fm}^{-1}$,
and a Gaussian-type potential between $\alpha$ particle and 
nucleon\cite{HagiKoi}:
$V_{\alpha {\rm N}}(r)=-v_0e^{-\beta r^2}$, $v_0=40.45\,{\rm MeV}$, $\beta=0.189\,{\rm fm}^{-2}$. For $^7_\Lambda$Li nucleus, one also has to add the 
$\Lambda$-nucleon potential to $V_{\rm cN}$, that is, 
$V_{\rm cN}=V_{\alpha N}+V_{\Lambda N}$. 
In order to construct it, we fold 
a $\Lambda$-nucleon potential $v_{\Lambda N}$ in the free space with the 
$\Lambda$ particle density in the $\alpha$ cluster, 
\begin{equation}
V_{\Lambda {\rm N}}(r)=
\int d\vec{r}'\rho_{\Lambda}(\vec{r}')v_{\Lambda {\rm N}}(\vec{r}-\vec{r}'). 
\end{equation}
For a Gaussian density distribution, 
$\rho_{\Lambda}(r)=(\pi b_\Lambda^2)^{-2/3}e^{-r^2/b_{\Lambda}^2}$ and a 
Gaussian $\Lambda$-nucleon potential, 
$v_{\rm \Lambda N}(r)=-v_0{\rm e}^{-r^2/b_v^2}$, 
this can be calculated analytically as 
\begin{equation}
V_{\Lambda {\rm N}}(r)
=
-v_0\left(\frac{b_v^2}{b_{\Lambda}^2+b_v^2}\right)^{3/2}
{\rm exp}\left[-\frac{r^2}{b_{\Lambda}^2+b_v^2}\right].
\end{equation}
We numerically solve the Schr\"{o}dinger equation with the 
potential $V(R)$, Eq. (\ref{pot}). 
The ground state is 
identified as the state with the node of 1 \cite{HagiKoi,merchant}. 
In this paper, we use the same width and strength parameters, 
$b_\Lambda, b_v$, and $v_0$ 
as those in Ref.\cite{HagiKoi}. 

\begin{figure}
\begin{center}
\includegraphics[scale=.5]{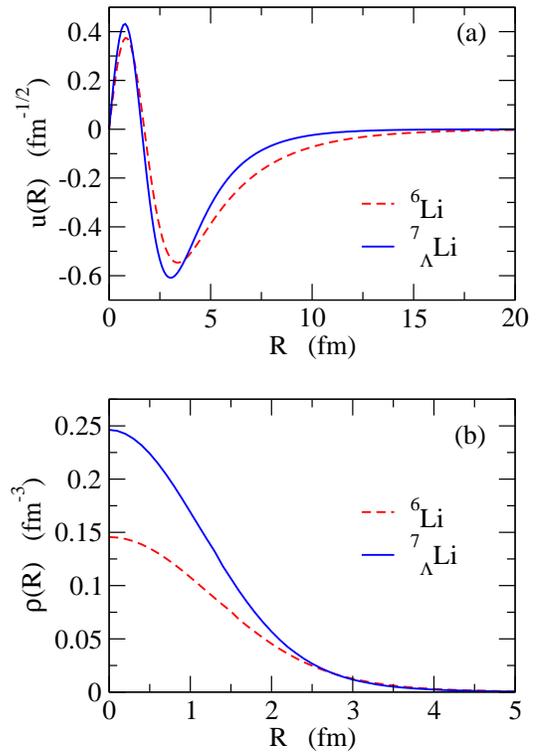}
\end{center}
\caption{(Color online) 
(upper panel) 
The intercluster radial wave functions $u(R)$, 
defined as $\psi_{\rm rel}(\vec{R})=\frac{u(R)}{R}Y_{00}(\hat{\vec{R}})$,  
for the ground state 
of $^6{\rm Li}$ (the dashed line) and $^7_{\Lambda}{\rm Li}$ (the solid line) 
nuclei 
obtained with the semi-microscopic cluster model.
The $\alpha$+deuteron and $^5_\Lambda$He + deuteron structures are 
assumed for 
$^6{\rm Li}$ and $^7_{\Lambda}{\rm Li}$, respectively. 
(the lower panel) 
The same as the upper panel, but for 
the one-body densities $\rho(R)$ 
calculated with $\psi_{\rm rel}$. }
\label{fig:density}
\end{figure}

Figure \ref{fig:density} shows the intercluster radial wave functions $u(R)$
so obtained, where $u(R)$ is defined with 
the intercluster wave function 
$\psi_{\rm rel}(\vec{R})$ as 
$\psi_{\rm rel}(\vec{R})=\frac{u(R)}{R}Y_{00}(\hat{\vec{R}})$. 
The dashed and solid lines 
are for $^6$Li and $^7_\Lambda$Li, respectively. 
The figure also shows the one-body densities $\rho(r)$ 
obtained as 
\begin{eqnarray}
\begin{aligned}
\rho(\vec{r})=\int d\vec{R}\,
|\psi_{\rm rel}(\vec{R})|^2
\left[ \rho_1(\vec{r}+a_1\vec{R})+\rho_2(\vec{r}-a_2\vec{R})\right],
\end{aligned}
\end{eqnarray}
where $a_1$ and $a_2$ are defined as $M_2/(M_1+M_2)$ and $M_1/(M_1+M_2)$, 
respectively, $M_1$ and $M_2$ being the mass of the cluster 1 and 2, 
respectively. 
We use Gaussian-type densities for $d$ and $\alpha$, 
$\rho_i(r)=A_i(\pi\gamma_i^2)^{-2/3}e^{-r^2/\gamma_i^2}$, where $A_i$ is the 
mass number of the cluster $i$ and $\gamma_i$= 
1.36 and 1.60 fm for $i=\alpha$ and $d$, respectively, 
while the density 
for $^5_\Lambda$He is given by $\rho_\alpha+\rho_\Lambda$.
One can see that 
the addition of the $\Lambda$ particle 
shifts 
the wave function of $^7_{\Lambda}{\rm Li}$ towards smaller distances as 
compared to the wave function of $^6$Li. 
The root mean square (rms) radii for $^6{\rm Li}$ and $^7_{\Lambda}{\rm Li}$
are 4.342 and 3.527 fm, respectively.
This amounts to a 
shrinkage of about 19\%, which 
agrees well with the experimental data \cite{tanida} as well 
as with more microscopic calculations 
in Refs.\cite{motoba,motoba2,hiyama}. 
These densities and the wave functions are used 
in the Glauber calculations for reaction cross sections. 

Let us now compute 
the reaction cross sections
for the $^6$Li+$^{12}$C and $^7_\Lambda$Li+$^{12}$C systems 
at $100 \,{\rm MeV/nucleon}$ with the Glauber theory. 
We first 
use the optical limit approximation (OLA) with a zero-range 
nucleon-nucleon interaction.  
The reaction cross section for $^6$Li in this approximation 
reads 
\begin{eqnarray}
\begin{aligned}
\sigma^{\rm (OLA)}_{\rm R}
=2\pi\int b\,db\,\left[ 1-{\rm e}^{-\sigma_{\rm NN}
\int d\vec{s}\,\rho_P^{(z)}(-\vec{b}+\vec{s})
\rho_{\rm T}^{(z)}(\vec{s})}\right],
\end{aligned}
\end{eqnarray}
where $\sigma_{\rm NN}$ is the total cross section for nucleon-nucleon 
scattering 
and $b$ is an impact parameter. 
$\rho^{(z)}$ is a $z$-integrated density, 
$\rho^{(z)}(\vec{s})\equiv\int dz\,\rho(\vec{r})$,
where $\vec{s}$ is 
the two-dimensional vector perpendicular to the beam axis, that is, 
$\vec{r}=(\vec{s},z)$. 
The reaction cross section for $^7_\Lambda$Li can also be obtained 
in a similar manner, by adding the contribution of the $\Lambda$ particle 
in the exponent (see Eq. (\ref{s}) below). 
In our calculations, we use the Gaussian density distribution for 
$^{12}$C with the width parameter of $\gamma_T=$1.89 fm. 

For comparison, we 
also compute the reaction cross sections 
using 
the few-body (FB) treatment\cite{tostevin}  
beyond the OLA, 
in order to take account of the cluster correlations in the projectiles 
(see also Ref. \cite{HoriSuzu} for an alternative method which 
is beyond the OLA). 
The reaction cross sections in the FB are given as 
\begin{eqnarray}
\begin{aligned}
\sigma^{\rm (FB)}_{\rm R}
=2\pi\int b\,db\,
\left[1-|\langle\psi_{\rm rel}|S_1(b_1)S_2(b_2)|\psi_{\rm rel}\rangle|^2\right], 
\end{aligned}
\end{eqnarray}
where 
$S_i(b_i)$ is the scattering matrix between the cluster $i$ in the 
projectile and the target nuclei evaluated in the OLA.  
For $d$ and $\alpha$, 
$S_i(b)$ is given by 
\begin{equation}
S_i(b)
={\rm exp}\left[-\frac{\sigma_{\rm NN}}{2}
\int d\vec{s}\,\rho_i^{(z)}(-\vec{b}+\vec{s})\rho_{\rm T}^{(z)}(\vec{s})\right].
\label{FB}
\end{equation}
Since we employ the Gaussian-type densities 
for all the clusters including the target,
it can be obtained analytically as 
\begin{eqnarray}
\begin{aligned}
S_i(b)
&={\rm exp}\left[-\frac{\sigma_{\rm NN}}{2}
\frac{A_iA_{\rm T}}{\pi(\gamma_i^2+\gamma_{\rm T}^2)}
{\rm e}^{-b^2/(\gamma_i^2+\gamma_{\rm T}^2)}\right]. 
\end{aligned}
\end{eqnarray}
For a hypernucleus $^5_{\rm \Lambda}{\rm He}$ 
Eq. (\ref{FB}) is extended as 
\begin{eqnarray}
\begin{aligned}
S_{^5_{\rm \Lambda}{\rm He}}(b)
={\rm exp}\bigl[&-\frac{\sigma_{\rm NN}}{2}
\int d\vec{s}\,\rho_\alpha^{(z)}(-\vec{b}+\vec{s})\rho_{\rm T}^{(z)}(\vec{s})
\\
&-\frac{\sigma_{\rm \Lambda N}}{2}
\int d\vec{s}\,\rho_\Lambda^{(z)}(-\vec{b}+\vec{s})\rho_{\rm T}^{(z)}(\vec{s})
\bigr],
\label{s}
\end{aligned}
\end{eqnarray}
where $\sigma_{\rm \Lambda N}$ is the total cross section 
for $\Lambda$-nucleon scattering.
We consider the energy region of 
$E\simeq 100\,{\rm MeV/nucleon}$, where 
experimental cross sections
for both $\sigma_{\rm NN}$ and $\sigma_{\rm \Lambda N}$ are available, although 
$\sigma_{\rm \Lambda N}$ has not been determined accurately. 
We take $\sigma_{\rm NN}=55.2$ mb\cite{suzuki} and
$\sigma_{\rm \Lambda N}$ in the range 
from 10 to 30 mb \cite{hauptman}. 

\begin{table}
\caption{The reaction cross sections $\sigma_{\rm R}$, in the units of mb, 
for $^6{\rm Li}$ and $^7_{\rm \Lambda}{\rm Li}$
incident on a $^{12}{\rm C}$ target at $100\,{\rm MeV/nucleon}$.
The values for $\Lambda$-nucleon scattering cross section \cite{hauptman}, 
$\sigma_{\rm \Lambda N}$, are given in the parentheses.
Both the results based on the few-body (FB) treatment and the optical 
limit approximation (OLA) for the Glauber theory are shown.
}
\begin{center}
\begin{tabular}{ccc}
\hline
\hline
\multicolumn{1}{c}{}&\multicolumn{1}{c}{FB}
&\multicolumn{1}{c}{OLA}\\
\hline
$^6{\rm Li}$&~~825.9&~~880.6\\
\hline
$^7_\Lambda{\rm Li}(\sigma_{\rm \Lambda N}=0\,{\rm mb})$&~~781.6&~~815.3\\
$(10\,{\rm mb})$&~~786.7&~~819.9\\
$(20\,{\rm mb})$&~~791.7&~~824.4\\
$(30\,{\rm mb})$&~~796.5&~~828.8\\
\hline
\hline
\end{tabular}
\end{center}
\label{tb:sigmar}
\end{table}

Table \ref{tb:sigmar} shows the reaction cross sections obtained 
with the FB and the OLA calculations for $^6{\rm Li}$ and 
$^7_{\rm \Lambda}{\rm Li}$. 
We notice that the OLA yields always larger reaction cross sections 
compared to the FB, indicating an importance of the cluster correlation 
in these nuclei. 
If one considers only the shrinkage of the nucleon distribution 
in $^7_\Lambda{\rm Li}$, 
neglecting the contribution of the $\Lambda$ particle to the reaction cross 
section (that is, setting $\sigma_{\rm \Lambda N}=0$), 
the reaction cross section for 
$^7_{\rm \Lambda}{\rm Li}$ decreases by 
about 5.4 \% in the FB and 7.4 \% in the OLA compared to 
the reaction cross section for $^6$Li. 
By including the $\Lambda$ particle contribution to the reaction 
cross section, the reduction is in the range of 4.7-3.6 \% in the FB and 
6.9-5.9 \% in the OLA. 
Although these values are somewhat smaller than the reduction in the 
rms radius, these results clearly indicate that 
the shrinkage effect in hypernuclei can be studied also 
with the reaction cross section. 

In summary, we have investigated reaction cross sections for $^6{\rm Li}$ and
$^7_{\rm \Lambda}{\rm Li}$ nuclei incident on a $^{12}{\rm C}$ target 
at $100\,{\rm MeV/nucleon}$. 
We have first applied for these projectile 
nuclei the semi-microscopic cluster model
in order to construct the ground state wave functions and the 
one-body densities. 
For a reaction theory, we have used 
the few-body (FB) treatment for the projectiles in the Glauber theory 
in order to 
take into account the cluster correlations. 
We have found that 
the reaction cross section for $^7_{\rm \Lambda}{\rm Li}$ is always smaller 
than that for $^6{\rm Li}$ by about a few percent. 
This fact suggests that the shrinkage effect induced by a $\Lambda$ particle 
in hypernuclei
can be studied with reaction cross sections as an alternative to 
$\gamma$-ray spectroscopy. 
It would be interesting if 
a measurement of reaction cross sections could be realized 
for hypernuclei in some future. 

\medskip

We thank H. Tamura, Y. Tanimura and Y. Urata for useful discussions. 
This work was supported by
JSPS KAKENHI Grant Number
22540262.

\end{document}